\documentclass[aps,prb,twocolumn,superscriptaddress]{revtex4-2}

\usepackage{graphicx}% Include figure files
\usepackage{dcolumn}% Align table columns on decimal point
\usepackage{bm}% bold math
\usepackage{graphicx}% Include figure files
\usepackage{derivative}
\usepackage{dcolumn}% Align table columns on decimal point
\usepackage{orcidlink}
\usepackage{bm}% bold math
\usepackage{hyperref}
\usepackage{braket}
\usepackage{amsmath}
\usepackage{xcolor}
\usepackage{amsfonts}

\begin{document}

\preprint{APS/123-QED}

\title{Information-Theoretic Signatures of Localization and Mobility Edges in Quasiperiodic Systems}

\author{Arpita Goswami\,\orcidlink{0009-0003-2241-7034}}
\affiliation{Department of Physics, Indian Institute of Technology Tirupati, India, 517619 \\
 %This line break forced with \textbackslash\textbackslash
}

\date{\today}

\begin{abstract}
We investigate localization transitions and mobility-edge phenomena in one-dimensional quasiperiodic lattice models using an information-theoretic framework based on the Tsallis entropy of single-particle eigenstates. Conventional diagnostics, such as the inverse participation ratio, characterize localization at the level of individual eigenstates but do not directly capture spectral heterogeneity arising from the coexistence of localized and extended states within the same spectrum. To address this, we employ the Tsallis entropy as a continuous, normalized functional of wavefunction amplitudes, where the entropic index $q$ provides a tunable sensitivity to different regions of the probability distribution, enhancing the contribution of localized peaks ($q>1$) or extended components ($q<1$). Building on this framework, we introduce an entropy-gradient susceptibility defined from the energy dependence of the Tsallis entropy, which probes variations in eigenstate structure across the spectrum. We show that this quantity clearly distinguishes global localization transitions from mobility-edge physics. In the Aubry--André model, where all eigenstates undergo a uniform transition, the entropy varies smoothly, resulting in a broad crossover in the susceptibility. In contrast, in systems hosting mobility edges—including a quasiperiodically modulated Su--Schrieffer--Heeger chain and the generalized Aubry--André model—the coexistence of localized and extended states produces sharp spectral variations, leading to a pronounced and system-size–independent peak. This peak provides a robust diagnostic of energy-resolved localization and enables the identification of characteristic scales associated with the mobility-edge regime. The qualitative behavior persists over a broad range of the entropic parameter $q$, with systematic variations reflecting its role as a tunable probe of spectral structure. Our results establish an information-theoretic approach that leverages the continuous $q$-dependence of Tsallis entropy to construct a derivative-based measure of spectral heterogeneity, providing a complementary and physically transparent diagnostic of mobility-edge phenomena beyond conventional state-resolved measures.
\end{abstract}

\maketitle

\section{Introduction}

In a low-dimensional disordered medium, Anderson localization \cite{Anderson_1} arises from destructive quantum interference, suppressing wave propagation and remaining a central problem in condensed matter physics. The paradigmatic Anderson localization scenario predicts the absence of diffusion in sufficiently disordered one-dimensional systems, while quasiperiodic lattices~\cite{QP_1, QP_2, QP_3, QP_4, QP_5, QP_6, QP_7, QP_8, QP_9, QP_10, QP_11} provide an alternative route to localization without true randomness. Among these, the Aubry–Andre (AA) model plays a special role due to its exact self-dual structure and sharp localization transition at a critical potential strength. Beyond such global localization transitions, a significant body of work has focused on systems exhibiting mobility edges (MEs)~\cite{ME_1, ME_2, ME_3, ME_4, ME_5, ME_6}, where localized and extended eigenstates coexist at different energies for a given disorder strength. Energy-resolved localization is particularly relevant for interacting systems, driven dynamics, and nonequilibrium phenomena. However, identifying mobility edges remains challenging, as commonly used diagnostics such as the inverse participation ratio (IPR)~\cite{ipr_1, ipr_2, ipr_3} or Lyapunov exponents~\cite{LE_1, LE_2} often require careful finite-size scaling and can be strongly influenced by rare states.

Information-theoretic measures provide an alternative framework for characterizing quantum states, offering a natural and basis-independent description of wave-function structure. In particular, entropy-based quantities capture the degree of spatial delocalization and have been widely employed to probe localization–delocalization transitions. Prominent examples include entanglement-entropy scaling studies \cite{entropic_measure_2,entropic_measure_3,entropy_measure_1,entropic_measure_4,entropic_measure_5,entropic_measure_6,entropic_measure_7, MISHRA2025, Sahoo2012} and observational entropy–based approaches \cite{observational_entropic}. In this work, we follow a complementary information-theoretic route \cite{inform_stat} based on Tsallis entropy \cite{Tsallis1988}, a well-established generalization of Boltzmann–Gibbs entropy~\cite{beale2021statistical} that has been successfully applied across a broad range of classical and quantum systems, including statistical mechanics \cite{Goswami_2025}, plasma physics \cite{TARUYA2003285, Jiulin2007, Plastino, DU2004262, PAVLOS2015113, campa}, quantum information and black-hole thermodynamics \cite{bl_h_1, ABE2001157, b_l_h2, CANOSA2005121}, as well as complex systems such as non-Markovian dynamics, fractal structures \cite{add_1, add_2, add_3}, and financial markets \cite{e22040452}.

The central motivation of this work is not to introduce an alternative measure of localization strength, but rather to directly probe spectral heterogeneity in quasiperiodic systems. Mobility edges are fundamentally collective spectral phenomena, characterized by the coexistence of localized and extended eigenstates within the same spectrum. While conventional diagnostics, such as the $IPR$, quantify the localization of individual eigenstates, the mean-$IPR$ can quantify this extent to some degree; however, it is extremely system-size-dependent and does not directly capture this coexistence. As a result, distinguishing global localization transitions from mobility-edge physics often requires indirect analysis or fine-tuned scaling procedures.

Here, we show that fluctuations of information-theoretic entropy across the energy spectrum provide a natural and robust diagnostic of spectral heterogeneity. Specifically, we introduce an entropy-gradient susceptibility, defined as the energy gradient of the normalized Tsallis entropy of single-particle eigenstates across the spectrum. Importantly, this susceptibility does not measure the localization strength of individual eigenstates; instead, it directly quantifies the coexistence of localized and extended states. In systems where all eigenstates localize collectively, such as the AA model, entropy fluctuations remain suppressed, yielding only a broad crossover near the transition. In contrast, systems hosting mobility edges—including a quasiperiodically modulated Su–Schrieffer–Heeger (SSH) chain with quasi-periodic onsite potential and the generalized Aubry–Andre (GAA) model—exhibit pronounced entropy fluctuations, resulting in a sharp, size-stable susceptibility peak that accurately tracks energy-resolved localization.

The Tsallis framework offers a particularly flexible setting for this analysis. The deformation parameter $q$ continuously tunes the entropy's sensitivity to rare, high-amplitude components of the wave function, allowing spectral heterogeneity to be probed systematically. While the special case $q=2$ admits a direct algebraic connection to the inverse participation ratio, we demonstrate that the mobility-edge signature persists across a broad range of entropic orders. This establishes that the susceptibility peak reflects genuine spectral structure rather than a moment-specific reformulation of participation-ratio–based measures. Recent efforts to diagnose mobility edges have employed a variety of approaches, including geometric diagnostics based on localization landscapes and percolation theory \cite{percolation}, spectral statistics such as level-spacing analysis in tilted lattices \cite{tilted_lattices}, and entanglement-entropy scaling in many-body systems \cite{entr_meas}. While powerful, these methods often suffer from finite-size sensitivity, basis dependence, or complexity associated with interacting systems. In contrast, the entropy-gradient susceptibility introduced here provides a $q$-robust (robust for $q \ge 1$, with systematic deviations for $q < 1$ due to enhanced rare-state weighting) and size-stable probe of spectral heterogeneity, enabling a sharp distinction between global localization transitions and mobility-edge scenarios across multiple quasiperiodic models. By reframing localization physics in terms of entropy fluctuations, our approach offers a complementary and broadly applicable perspective that bridges statistical mechanics, quantum information, and condensed matter physics.

The paper is organized as follows. In Sec.~\ref{sec_1} we describe the model systems considered for this study. In Sec.~\ref{sec_2} we introduce the Tsallis entropy and define the entropy-gradient susceptibility employed in this work. Numerical results are presented in Sec.~\ref{sec_3}, followed by analytical insights in Sec.~\ref{sec_4}. Finally, we summarize our conclusions in Sec.~\ref{sec_5}.

\section{Models} \label{sec_1}

In this section, we introduce the quasiperiodic lattice models studied in this work. These models are chosen to represent distinct localization scenarios:

(i) \textbf{Aubry--Andr\'e (AA) Model:} This model is known to have a global localization transition affecting the entire spectrum at a particular critical strength of onsite potential ($\lambda_c$).

(ii) \textbf{Dimerized Su--Schrieffer--Heeger (SSH) chain subject to a quasiperiodic onsite potential (SSH) and Generalized Aubry--Andr\'e (GAA) Model:} These two systems are known to exhibit mobility edges, where localized and extended eigenstates coexist between two fixed disorder strengths.

Analysis of these models allows us to directly assess the ability of entropic observables to distinguish between homogeneous and energy-resolved localization in different one-dimensional quasi-periodic systems.
\begin{figure}
    \centering
    \includegraphics[width=0.48\textwidth]{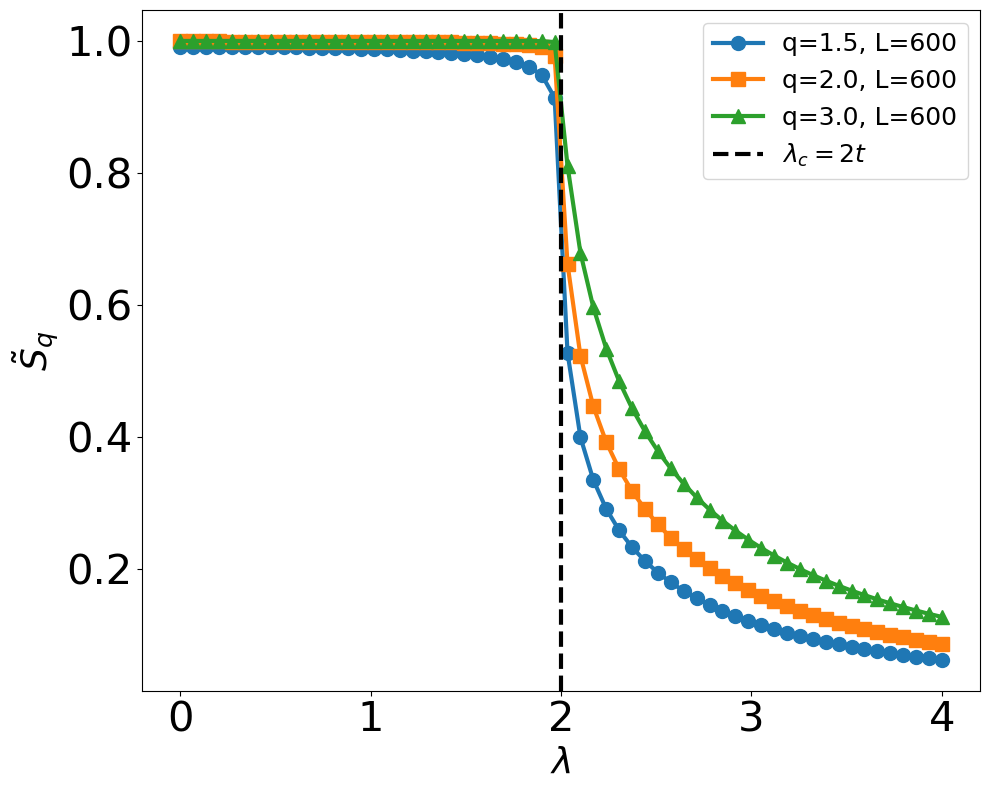}
    \caption{Normalized Tsallis entropy $\tilde{S_q}$ as a function of quasiperiodic potential strength $\lambda$ for the Aubry–André (AA) model at different entropic orders $q$, with system size $L=600$. The entropy decreases monotonically with increasing $\lambda$, reflecting the simultaneous localization of all eigenstates at the critical point $\lambda_c=
2t$}
    \label{plt_1}
\end{figure}

\begin{figure}[t]
\includegraphics[width=0.48\textwidth]{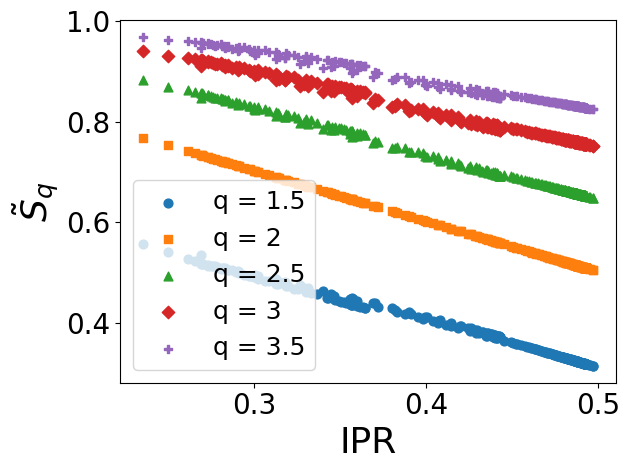}
    \caption{
    Correlation between normalized Tsallis entropy $\tilde S_q$ and inverse participation ratio (IPR) for a representative parameter set in a SSH Hamiltonian for different q values.
    States with large IPR correspond to low entropy, confirming consistency between entropic and conventional localization measures.
    }
    \label{plt_2}
\end{figure}

\subsection{Aubry--Andr\'e (AA) Model}

We begin with the one-dimensional Aubry--Andr\'e (AA) model, described by the tight-binding Hamiltonian
\begin{equation} \label{H_AA}
H_{\mathrm{AA}} = -t \sum_{n} \left( c_{n+1}^\dagger c_n + \text{h.c.} \right)
+ \lambda \sum_n \cos(2\pi \beta n + \phi)\, c_n^\dagger c_n ,
\end{equation}
where $t$ is the nearest-neighbor hopping amplitude, $\lambda$ controls the strength of the quasiperiodic potential, $\beta$ is an irrational number, and $\phi$ is a phase offset. $c_n$ and $c_n^\dagger$ are the fermionic annihilation and creation operators, respectively. Throughout this work, we fix $t=1$ and choose $\beta = (\sqrt{5}-1)/2$.

\begin{figure*}[t]
    \centering
    \includegraphics[width=1\textwidth]{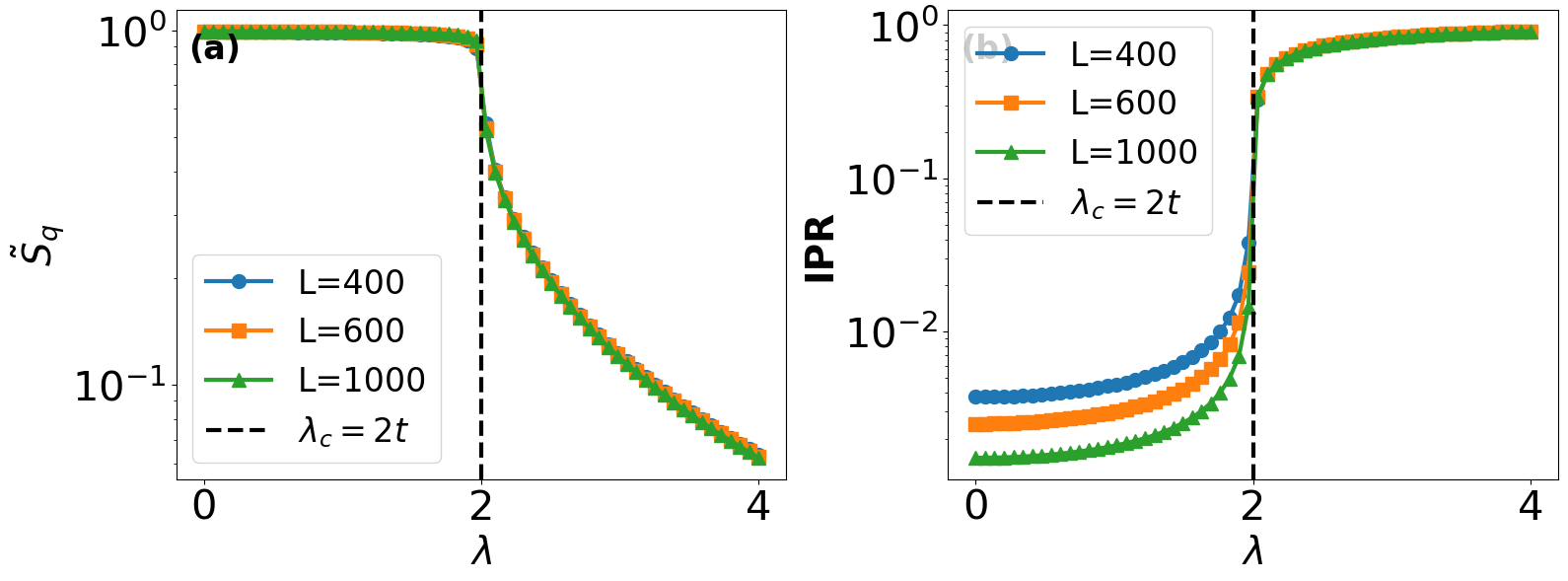}
    \caption{The scaling behaviour of (a) $\tilde{S_q}$ at $q=1.5$ and (b) IPR as a function of $\lambda$ for different $L$ for AA model, where in the case of $\tilde{S_q}$ scaling the $\lambda_c$ is denoted as the point where $\tilde{S_q}$ starts decreasing from maximum value $1$ and for $IPR$ scaling $\lambda_c$ is denoted as the point where the $IPR$ datas for different $L$ first collapsed.}
    \label{plt_3}
\end{figure*}

The AA model is self-dual and exhibits a sharp localization transition at $\lambda_c = 2t$. For $\lambda < \lambda_c$, all eigenstates are extended, while for $\lambda > \lambda_c$, all eigenstates become exponentially localized. Importantly, the transition occurs simultaneously for the entire spectrum, and the model does not host a mobility edge. As such, the AA model serves as a benchmark case for a spectrally homogeneous localization transition.
\begin{figure}
    \centering
    \includegraphics[width=0.48\textwidth]{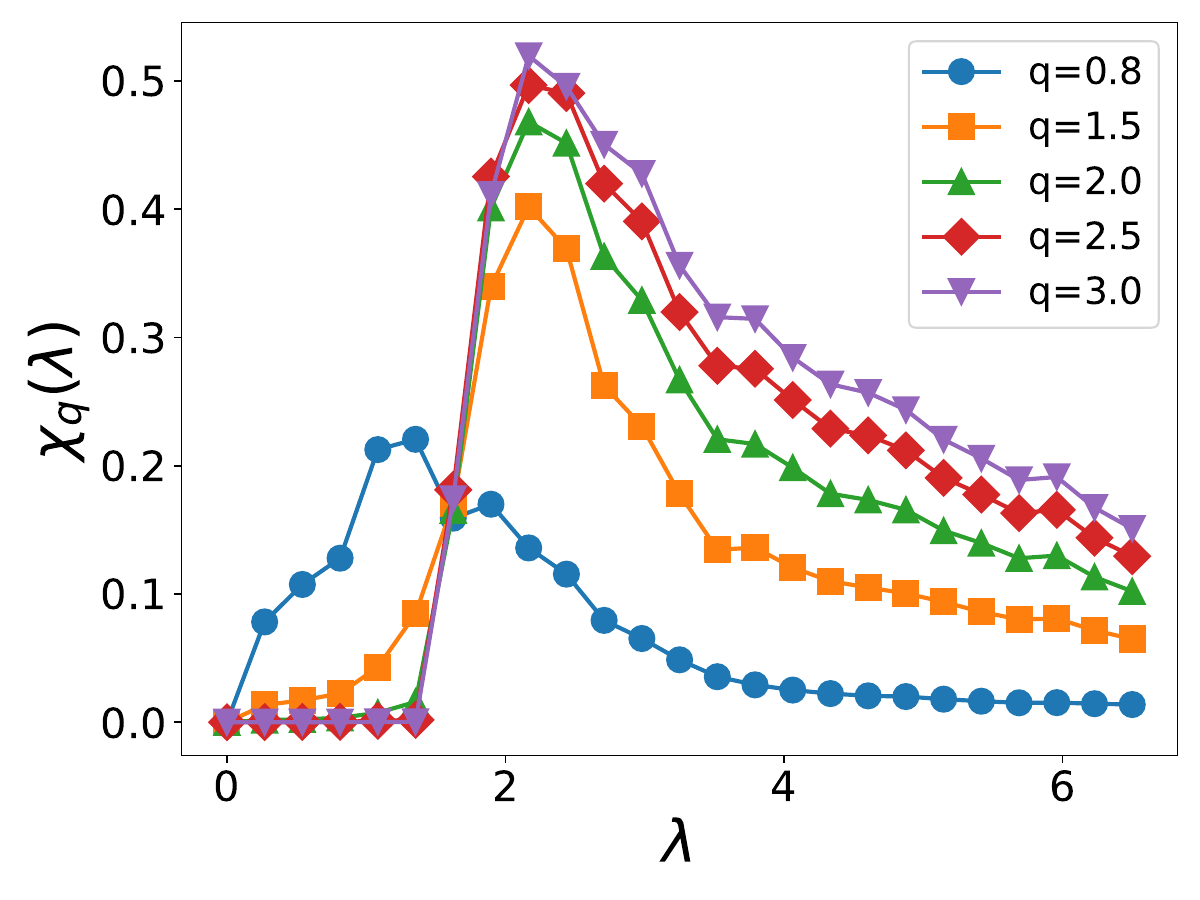}
    \caption{
Entropy-gradient susceptibility $\chi_q(\lambda)$ of the GAA model as a function of the quasiperiodic potential strength $\lambda$ for $q=1.5,2,2.5,3$.
For all entropic orders, the susceptibility develops a pronounced peak at intermediate coupling, signaling the coexistence of localized and extended eigenstates associated with a mobility edge.
While the peak height and width depend systematically on $q$, its position remains nearly unchanged, demonstrating that the mobility-edge signature is robust for $q \ge 1$, with systematic deviations for $q < 1$ due to enhanced rare-state weighting.
}
\label{plt_4}

\end{figure}

\begin{figure}
    \centering
    \includegraphics[width=0.48\textwidth]{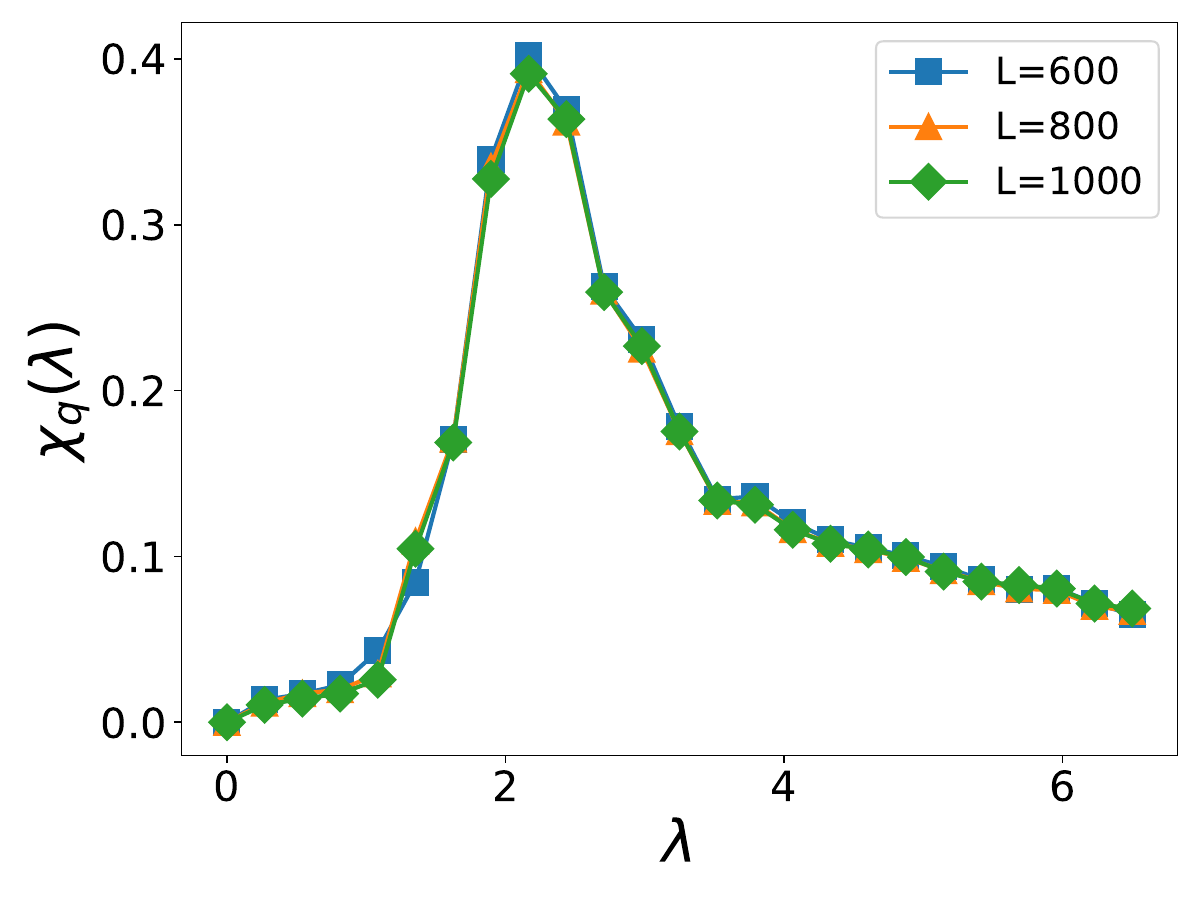}
    \caption{
Entropy-gradient susceptibility $\chi_q(\lambda)$ of the GAA model as a function of the quasiperiodic potential strength $\lambda$ for $q=1.5$ for $L=600, 800, 1000$. The peak height and width do not depend on $L$; its position also remains unchanged, demonstrating that the mobility-edge signature is robust against system size.
}
\label{plt_ex1}

\end{figure}

\subsection{SSH Model with Quasiperiodic Potential}

To investigate mobility-edge physics arising from competing lattice effects, we next consider a dimerized Su--Schrieffer--Heeger (SSH) chain subject to a quasiperiodic onsite modulation. The Hamiltonian is given by
\begin{equation} \label{H_SSH}
\begin{split}
H_{\mathrm{SSH}} = &
\sum_n \Big[
t_1\, c_{n,A}^\dagger c_{n,B}
+ t_2\, c_{n,B}^\dagger c_{n+1,A}
+ \text{h.c.}
\Big] \\ 
&+ \sum_{n,\alpha} \lambda \cos(2\pi \beta n + \phi)\,
c_{n,\alpha}^\dagger c_{n,\alpha},
\end{split}
\end{equation}
where $t_1$ and $t_2$ denote the intra- and intercell hopping amplitudes, respectively, and $\alpha = A,B$ labels the sublattices. For all numerical calculations, we fix $t_1 = 0.6$ and $t_2 = 0.4$.

\begin{figure}[t]
    \centering
    \includegraphics[width=0.48\textwidth]{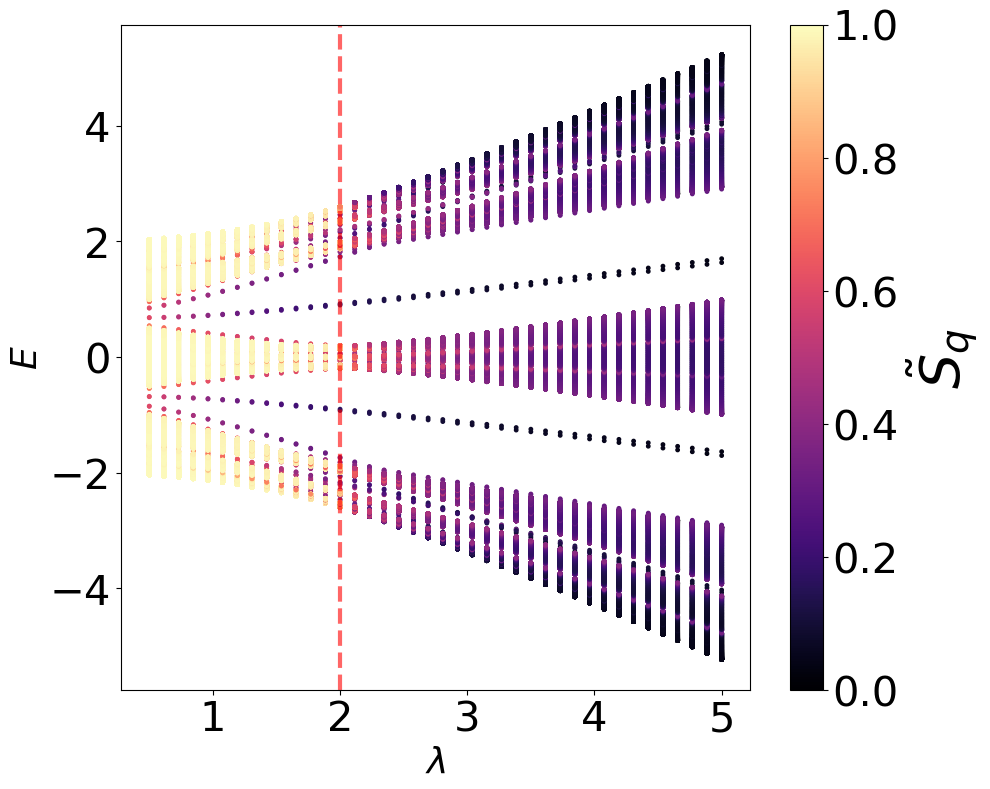}
    \caption{
    Normalized Tsallis entropy $\tilde S_q$ of the AA model as a function of quasiperiodic potential strength $\lambda$ at $q=1.5$ and $L=600$.
    The entropy decreases monotonically, reflecting the simultaneous localization of all eigenstates at the critical point $\lambda_c = 2t$.
    }
    \label{plt_5}
\end{figure} 
\begin{figure}[t]
    \centering
    \includegraphics[width=0.48\textwidth]{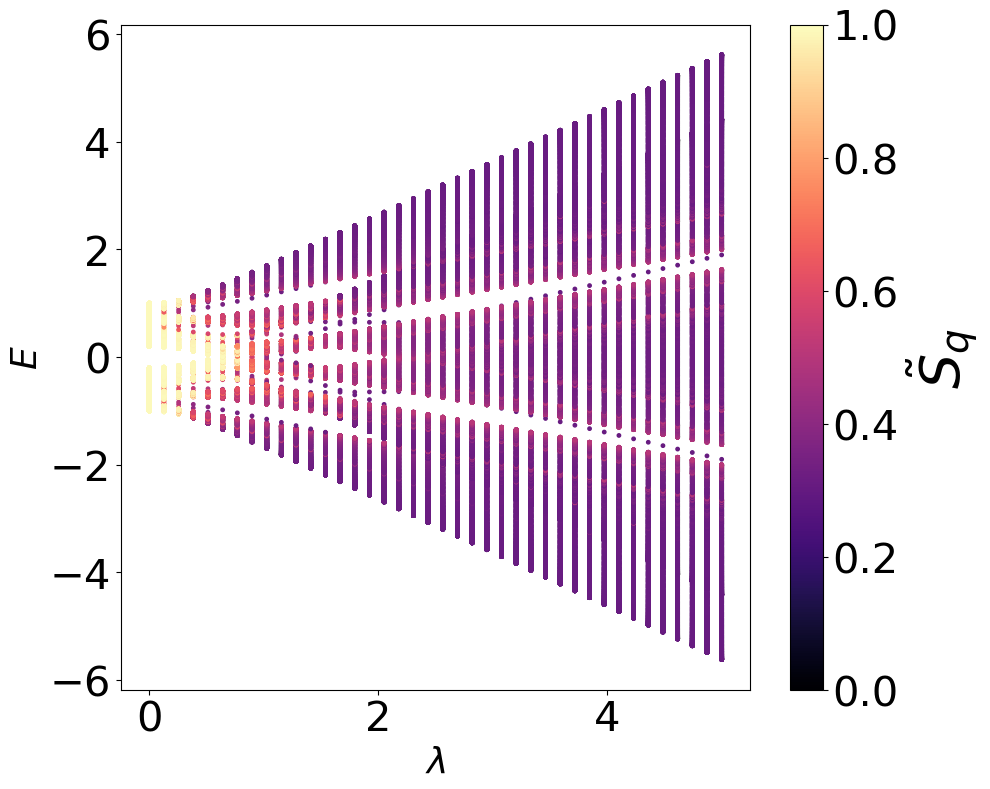}
    \caption{
    Scatter plot of normalized Tsallis entropy $\tilde S_q$ for the SSH model with a quasiperiodic onsite potential.
    Color encodes entropy magnitude, with darker regions corresponding to localized states.
    The curved boundary separating localized and extended states indicates the presence of a mobility edge. Calculations are done for $q=1.5$ and $L=600$.
    }
    \label{plt_6}
\end{figure}

\begin{figure}[t]
    \centering
    \includegraphics[width=0.48\textwidth]{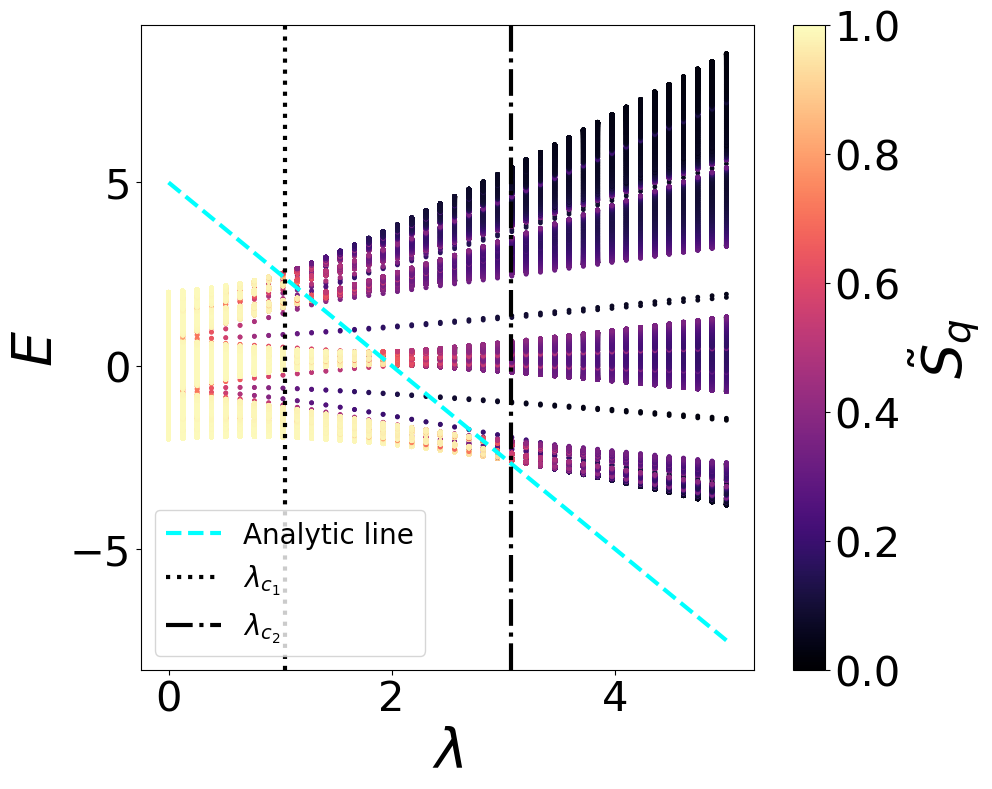}
    \caption{
    Normalized Tsallis entropy $\tilde S_q$ of the GAA model as a function of quasiperiodic potential strength $\lambda$ at $q=1.5$ and $L=600$.
    The entropy behaviour shows the presence of MEs. }
    \label{plt_7}
\end{figure}

\begin{figure}[t]
    \centering
    \includegraphics[width=0.48\textwidth]{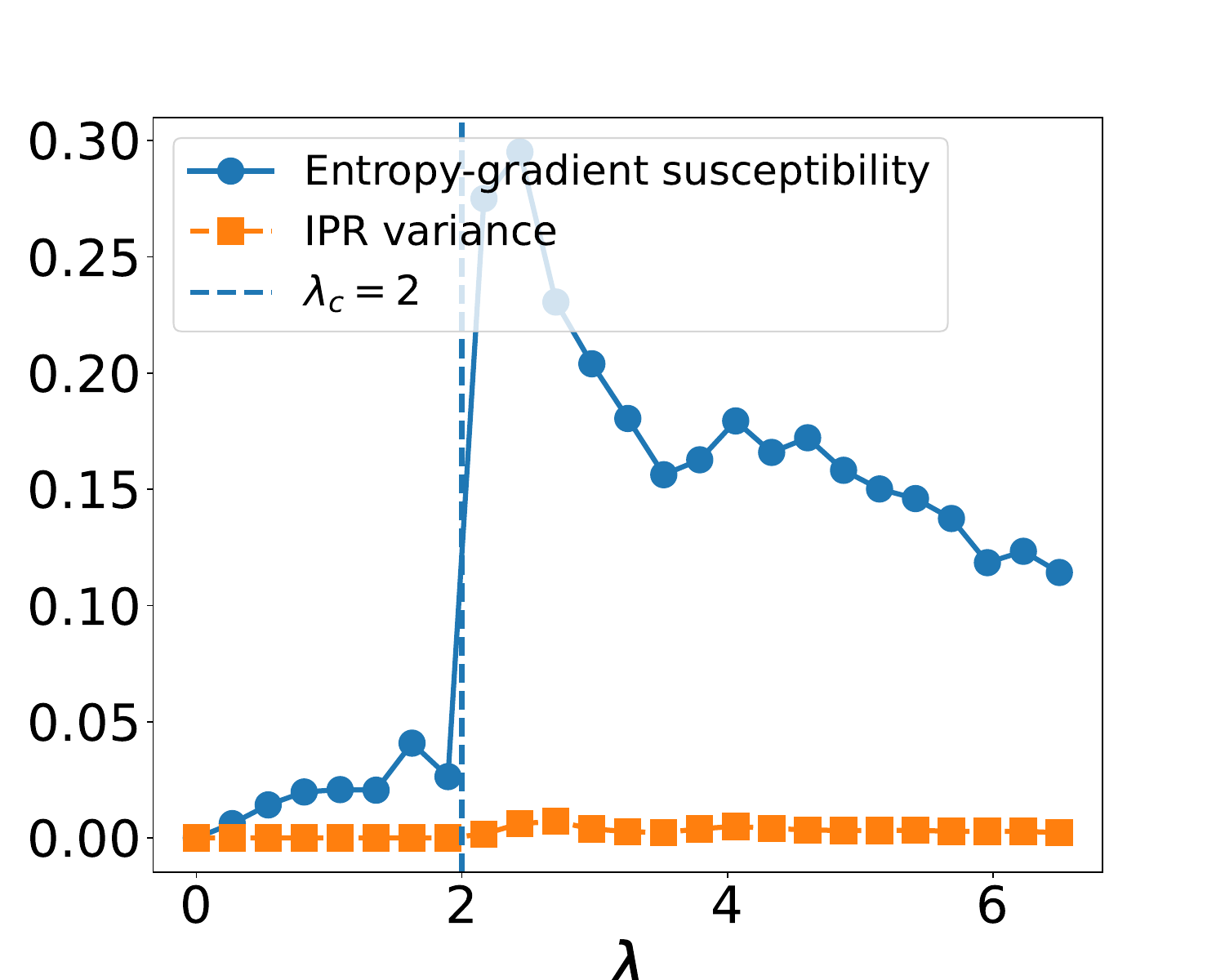}
    \caption{
    Entropy-gradient susceptibility $\chi_q(\lambda)$ of the AA model for different system sizes.
    The absence of a pronounced peak reflects the lack of energy-resolved localization, consistent with a global localization transition.
    }
    \label{plt_8}
\end{figure}

\begin{figure}[t]
    \centering
    \includegraphics[width=0.48\textwidth]{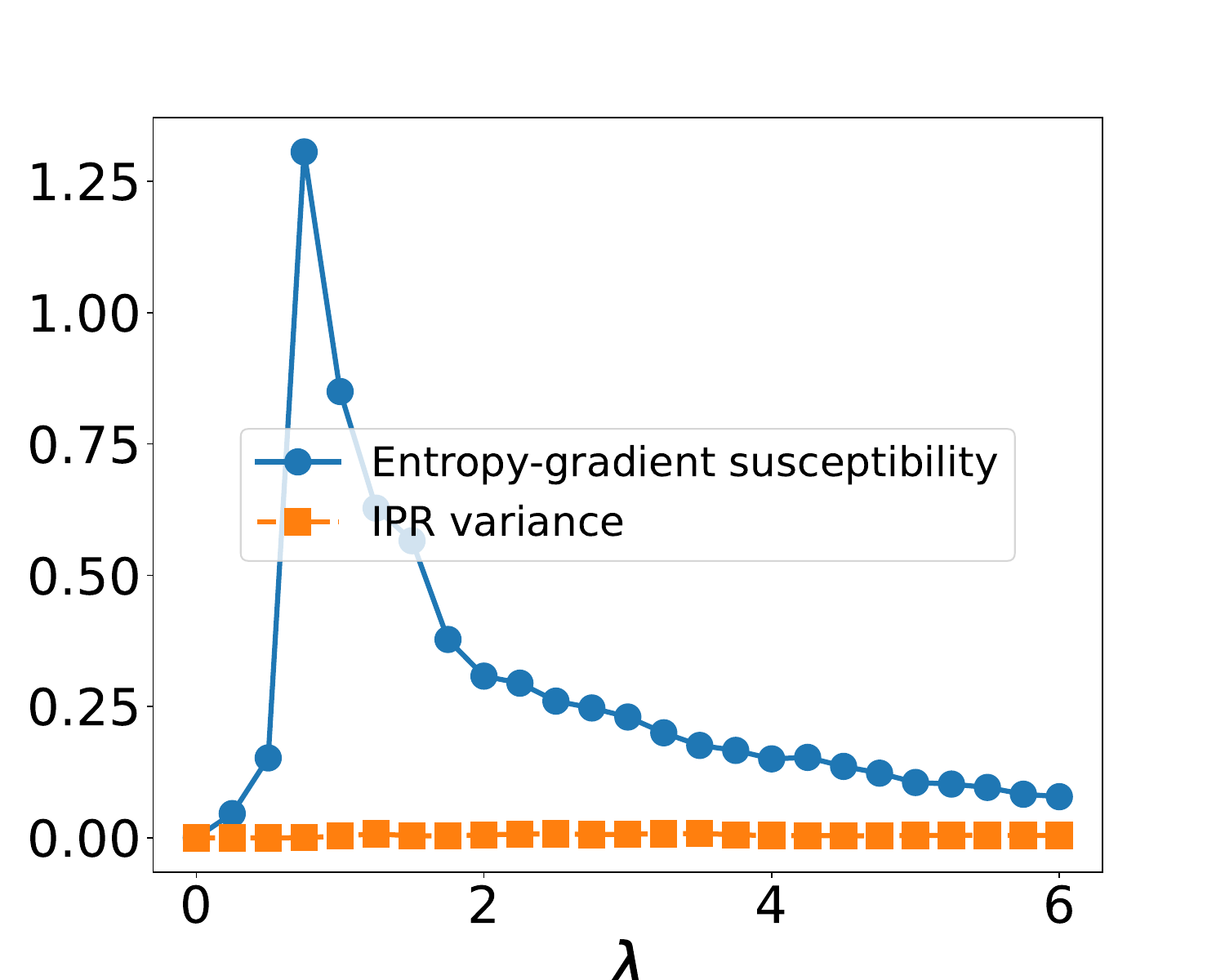}
    \caption{
    Entropy-gradient susceptibility $\chi_q(\lambda)$ of the SSH model for different system sizes.
    A pronounced peak emerges and sharpens with increasing system size, reflecting enhanced entropy fluctuations due to the coexistence of localized and extended states.
    }
    \label{plt_9}
\end{figure}

\begin{figure}[t]
    \centering
    \includegraphics[width=0.48\textwidth]{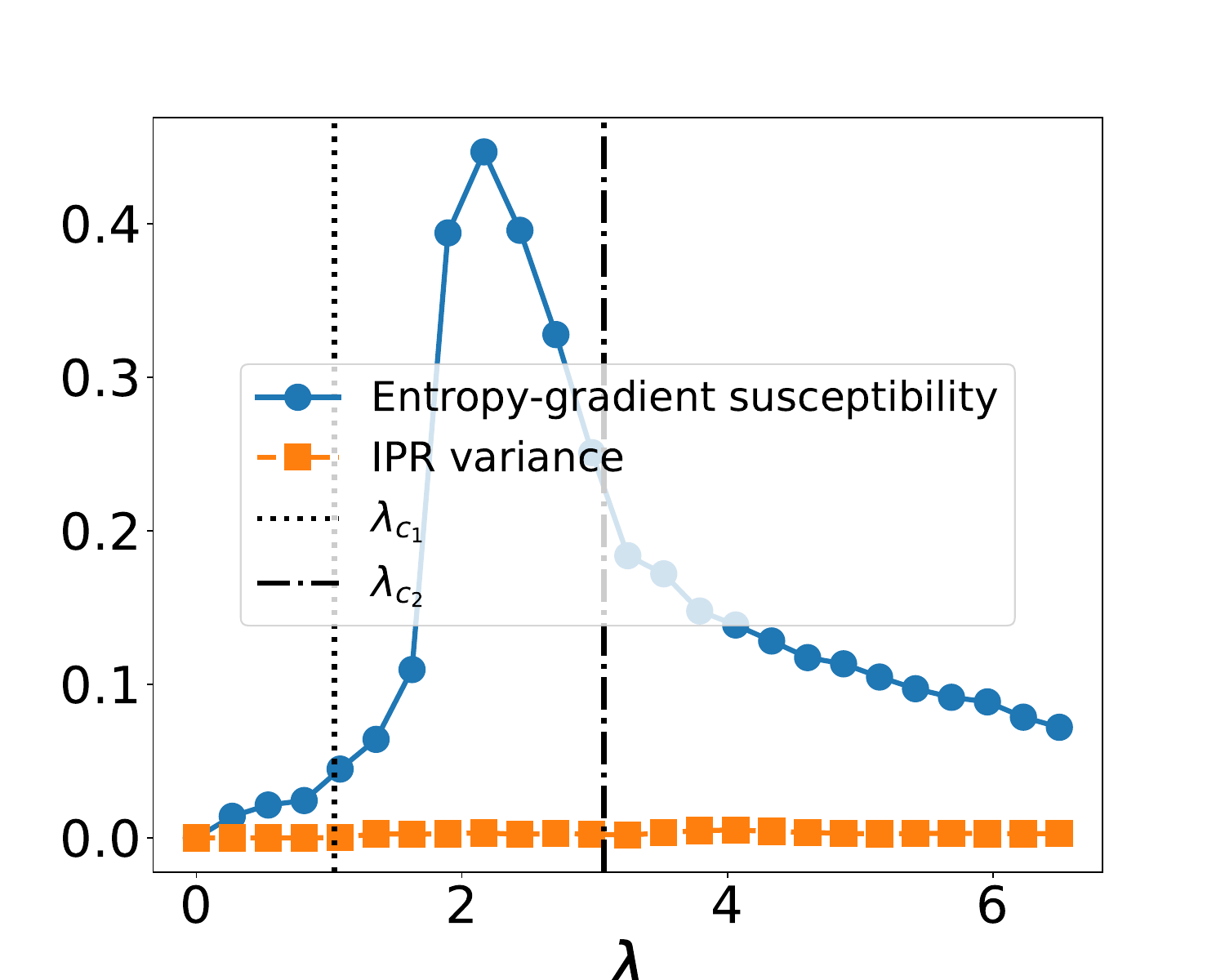}
    \caption{
   Entropy-gradient susceptibility $\chi_q(\lambda)$ of the GAA model for different system sizes. A pronounced and systematically sharpening peak emerges at intermediate quasiperiodic strength, signaling enhanced spectral heterogeneity due to the coexistence of localized and extended eigenstates.
    }
    \label{plt_10}
\end{figure}

\subsection{Generalized Aubry--Andr\'e Model}

Finally, we consider the one-dimensional generalized Aubry--Andr\'e (GAA) model, which extends the AA model by introducing a tunable parameter that generates an exact mobility edge. The Hamiltonian is given by
\begin{align} \label{H_GAA}
H_{\mathrm{GAA}} =& -t \sum_{n} \left( c_{n+1}^\dagger c_n + \text{h.c.} \right)\\ \notag
&+ \sum_n
\frac{\lambda \cos(2\pi \beta n + \phi)}
{1 - \alpha \cos(2\pi \beta n + \phi)}\,
c_n^\dagger c_n ,
\end{align}

where $\alpha \in (0,1]$ controls the deviation from the AA limit, recovered at $\alpha=0$. In our numerical calculations, we fix $\alpha = 0.4$.

For $\alpha \neq 0$, the model obeys an energy-dependent duality symmetry that leads to an exact mobility edge separating localized and extended eigenstates. The mobility-edge condition is known to be as follows:
\begin{equation} \label{ME_GAA}
\alpha E = 2t - \lambda.
\end{equation}
This closed-form relation allows for direct comparison between entropic diagnostics and analytically known localization boundaries, providing a straightforward test of the entropy-gradient susceptibility as a probe of energy-resolved localization.

\begin{figure*}[t]
    \centering
    \includegraphics[width=0.85\linewidth]{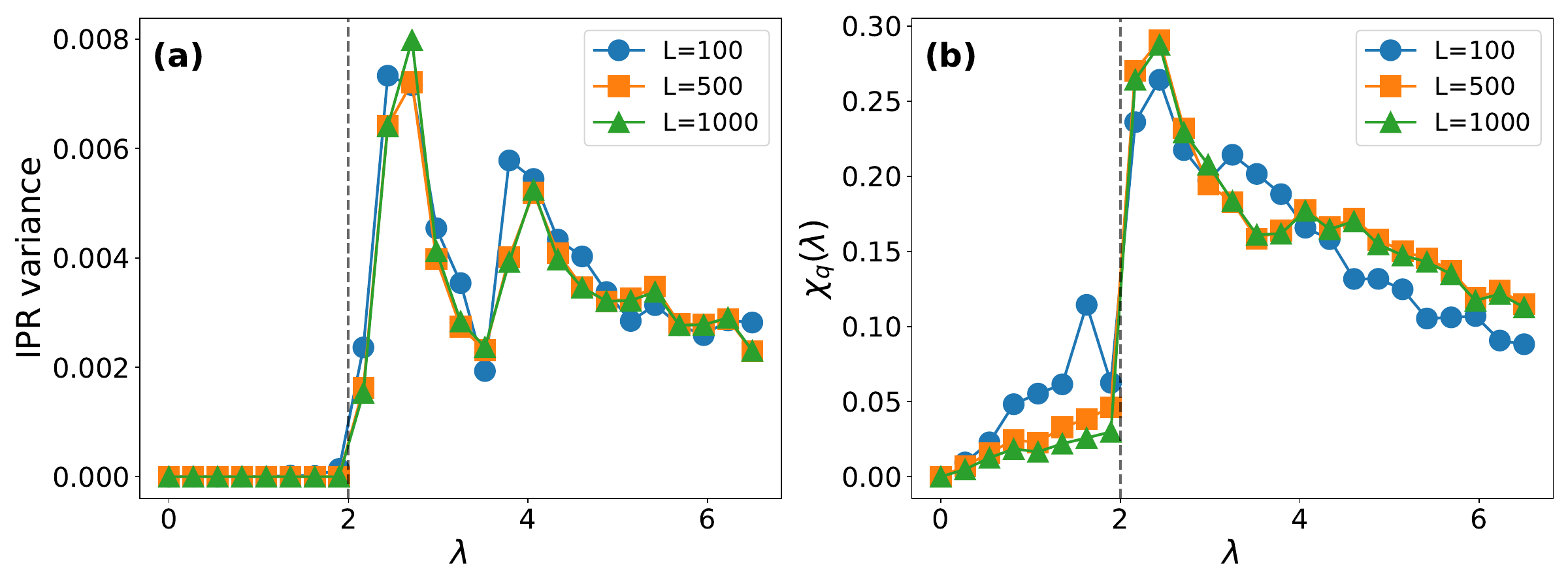}
    \caption{(a) Variance of the inverse participation ratio (IPR) as a function of the quasiperiodic potential strength $\lambda$ for the AA model. The variance is highly sensitive to fluctuations in individual eigenstates. (b) Spectral entropy-gradient measure as a function of $\lambda$ for the same model. In contrast to the IPR variance, this quantity exhibits a smooth crossover without any pronounced peak, consistent with the absence of spectral coexistence in the AA model, where all eigenstates undergo a collective localization transition at the critical point.}
    \label{plt_11}
\end{figure*}

\begin{figure*}[t]
    \centering
    \includegraphics[width=0.85\linewidth]{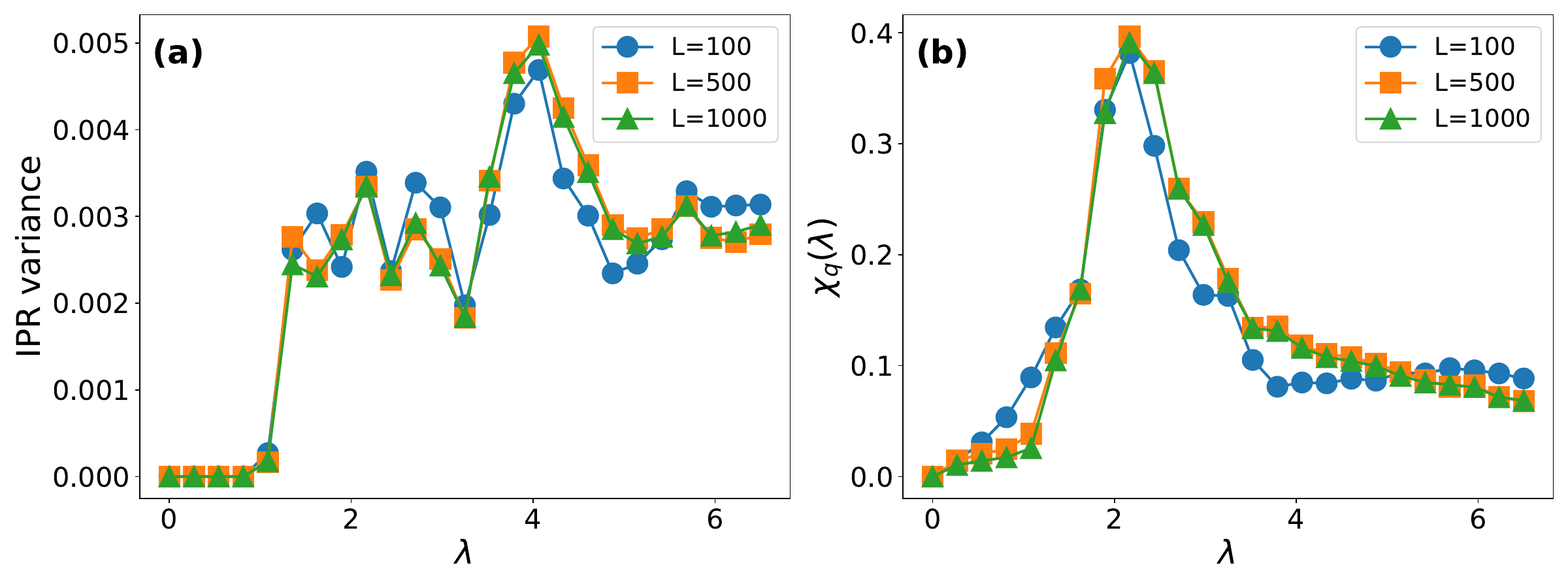}
    \caption{(a) Variance of the inverse participation ratio (IPR) as a function of the quasiperiodic potential strength $\lambda$ for the GAA model. The variance shows enhanced fluctuations in the intermediate regime, reflecting the presence of both localized and extended states; (b) Spectral entropy-gradient measure as a function of $\lambda$ for the same model. In contrast, a pronounced and sharply defined peak emerges in the intermediate regime, signaling the coexistence of localized and extended eigenstates characteristic of mobility-edge systems. The peak position remains stable across different system sizes, indicating that the entropy-based measure provides a robust and size-insensitive diagnostic of spectral heterogeneity.}
    \label{plt_12}
\end{figure*}

\section{Tsallis Entropy and Entropy-gradient susceptibility} \label{sec_2}

In this section, we introduce the entropic framework for characterizing the localization properties of single-particle eigenstates in quasiperiodic systems. We begin with the definition of the Boltzmann--Gibbs--Shannon entropy~\cite{beale2021statistical} and its nonextensive generalization, the Tsallis entropy~\cite{Tsallis1988}, and then introduce the entropy-gradient susceptibility that serves as the primary diagnostic throughout this work.

For a discrete probability distribution
$P=\{p_i \mid i=1,2,\dots,N\}$,
the Shannon entropy is defined as
\begin{equation}
S_{\mathrm{BG}} = -k_B \sum_{i=1}^{N} p_i \ln p_i,
\end{equation}
where $k_B$ is the Boltzmann constant. Tsallis proposed a one-parameter generalization of this entropy, given by
\begin{equation}
S_q = k_B \frac{1 - \sum_i p_i^q}{q - 1},
\quad q \in \mathbb{R},
\end{equation}
which reduces to $S_{\mathrm{BG}}$ in the limit $q \to 1$. The parameter $q$ controls the sensitivity of the entropy to fluctuations in the underlying probability distribution and has been widely used in the study of complex and nonergodic systems~\cite{bhattacharjee2026, 6zfg-q1b9, Han2025}.

In the present work, the probability distribution is constructed from the spatial profile of single-particle eigenstates. For a normalized eigenstate $\psi^{(n)}$ defined on $N$ lattice sites, we define
\begin{equation}
p_i^{(n)} = |\psi_i^{(n)}|^2.
\end{equation}
The Tsallis entropy of order $q$ for the $n$th eigenstate is then
\begin{equation}
S_q^{(n)} = \frac{1 - \sum_i \left(p_i^{(n)}\right)^q}{q - 1}.
\end{equation}
The above expression is closely related to generalized participation ratios $P_q = \sum_i |\psi_i|^{2q}$, which are widely used to characterize localization and multifractality~\cite{frac_1, frac_2, frac_3}. While $P_q$ can in principle be defined for real $q$, in practice it is typically evaluated at discrete integer values, thereby providing only a coarse sampling of wavefunction moments. In contrast, the Tsallis entropy furnishes a continuous and normalized functional of $q$, enabling a systematic and smooth interpolation across different moment orders. This continuous $q$-dependence allows one to tune the sensitivity of the entropy to rare, high-amplitude components of the wavefunction in a controlled manner. Importantly, the resulting smooth entropy landscape is essential for constructing stable energy-resolved derivatives, which serve as the basis for the entropy-gradient susceptibility introduced below. This enables a direct probe of spectral heterogeneity that is not naturally accessible in discrete-moment-based approaches.
\par The entropic index $q$ controls the statistical weight assigned to different regions of the wavefunction amplitude distribution. For $q > 1$, the entropy becomes more sensitive to high-probability components, thereby enhancing the contribution of localized peaks. In contrast, for $q < 1$, the entropy assigns relatively larger weight to low-probability components, amplifying the contribution of rare configurations corresponding to the extended tails of the wavefunction. This tunable weighting of typical versus rare states provides a controlled way to probe different structural features of eigenstates, particularly relevant in quasiperiodic systems that exhibit a coexistence of localized and extended states.
\par
To enable comparison across different system sizes, we introduce the normalized entropy
\begin{equation}
\tilde S_q^{(n)} =
\frac{S_q^{(n)}}{S_q^{\mathrm{max}}},
\quad
S_q^{\mathrm{max}} = \frac{1 - N^{1-q}}{q - 1},
\end{equation}
where $S_q^{\mathrm{max}}$ corresponds to a completely delocalized state with $p_i = 1/N$. Extended eigenstates thus yield $\tilde S_q^{(n)} \simeq 1$, while localized states produce significantly smaller values. It is to be noted that $S_q^{(n)}$ provides an entropic reformulation of wavefunction localization, with extended states characterized by $S_q^{(n)} \approx 1$ and strongly localized states by $S_q^{(n)} \approx 0$. $IPR$ for $n^{th}$ eigenstate of the spectra $\psi_n$ can be defined as follows:
\begin{equation}
    IPR^{(n)}=\sum_{i=1}^{N} |\psi^{(n)}_i|^4,
\end{equation}
where the sum $i$ goes over the lattice sites, and $N$ is the total number of lattice sites. The explicit relationship between $IPR$ and $S_q$ for different q-values has been demonstrated in the later sections.

While the entropy of individual eigenstates already captures localization at the wavefunction level, the central observable of this work is the fluctuation of entropy across the spectrum, which is defined as the entropy gradient susceptibility. We define the \emph{entropy-gradient susceptibility}
\begin{equation}
\label{susceptibility}
\chi_q =
\left\langle
\left|
\frac{d \tilde S_q(E)}{dE}
\right|
\right\rangle_E ,
\end{equation}
which measures the variation of the normalized Tsallis entropy over all eigenstates at a given quasiperiodic potential strength $\lambda$.

The physical significance of $\chi_q$ lies in its sensitivity to spectral heterogeneity. In systems exhibiting a global localization transition, such as the AA model, all eigenstates localize simultaneously, resulting in uniformly suppressed entropy and a weak susceptibility. In contrast, systems hosting a mobility edge display the coexistence of localized and extended states at fixed parameter values, leading to enhanced entropy fluctuations and a pronounced peak in $\chi_q$.

Although the Tsallis entropy at $q=2$ is algebraically related to the inverse participation ratio, the entropy-gradient susceptibility probes a fundamentally different aspect of localization physics. Unlike $IPR$-based diagnostics, which are strongly influenced by rare states and exhibit significant finite-size effects, the susceptibility captures collective spectral structure and remains robust across system sizes. This makes $\chi_q$ a powerful and reliable diagnostic for distinguishing global localization transitions from mobility-edge phenomena in quasiperiodic systems.
\par

For each Hamiltonian, we diagonalize the system and compute the normalized Tsallis entropy $\tilde S_q(E)$ for all eigenstates. To construct the entropy-gradient susceptibility, the spectral data are coarse-grained by partitioning the energy spectrum into 150 uniform bins, with each bin containing multiple states to ensure statistical stability. The resulting entropy profile is smoothed using a moving-average procedure to suppress residual fluctuations without affecting global trends. Energy derivatives are evaluated with a small cutoff to avoid numerical instabilities associated with near-degenerate levels. To mitigate the influence of rare states and outliers, a median-absolute-deviation (MAD) filtering scheme is applied. The susceptibility $\chi_q(\lambda)$ is then defined as the mean absolute value of the filtered entropy gradient across the spectrum. We have verified that the qualitative behavior and peak structure of $\chi_q$ are robust against variations in binning, smoothing, and filtering parameters, indicating that the observed features reflect intrinsic spectral properties rather than numerical artifacts. This procedure yields a stable, system-size-insensitive diagnostic of spectral heterogeneity, enabling clear identification of mobility-edge signatures. We have further confirmed that the extracted peak positions remain unchanged within numerical uncertainty under these variations.

\section{Numerical Results}
\label{sec_3}

In this section, we investigate localization properties using the entropy-based measures introduced in Sec.~\ref{sec_2} for all three quasiperiodic models with distinct localization behavior. Figure~\ref{plt_1} shows the normalized Tsallis entropy ($\tilde{S_q}$) as a function of $\lambda$ for different $q$ values in the AA model. For $\lambda < 2t$ (delocalized regime), $\tilde{S_q}$ attains its maximum value of unity for all $q$. In contrast, for $\lambda > 2t$ (localized regime), smaller $q$ values yield significantly reduced entropy, reflecting the subextensive nature of entropy in localized states. This demonstrates that varying $q$ enhances sensitivity to localization properties. Figure~\ref{plt_2} presents $\tilde{S_q}$ versus $IPR$ for the SSH model. As $q$ increases, the correlation between $\tilde{S_q}$ and $IPR$ weakens and eventually becomes nearly invariant.

\subsection{Scaling Behavior of Entropy and IPR}
\label{sec:scaling_AA_GAA}

To highlight the advantage of generalized entropy over the conventional inverse participation ratio (IPR), we compare their scaling behavior in the AA model for system sizes $L=400, 600, 1000$ at $q=1.5$. Figure~\ref{plt_3}(a) shows that $\tilde{S_q}$ remains system-size independent across both localized and delocalized regimes. In contrast, Fig.~\ref{plt_3}(b) demonstrates that $IPR$ exhibits strong size dependence in the delocalized regime ($\lambda < 2t$). This motivates the use of entropy-based diagnostics as complementary probes of localization transitions in finite-size systems.

\subsection{Dependence on the Tsallis Deformation Parameter}
\label{sec:q_dependence}
The entropy-gradient susceptibility is well defined for arbitrary values of the Tsallis deformation parameter $q$. As shown in Fig.~\ref{plt_4}, $\chi_q(\lambda)$ in the GAA model develops a pronounced peak at intermediate coupling for all $q$, signaling the coexistence of localized and extended eigenstates. For $q \gtrsim 1$, the peak position remains nearly invariant, confirming that the mobility-edge signature is robust and not tied to a specific entropic order. However, for $q < 1$, a slight shift in the peak position is observed. This can be attributed to the enhanced weighting of low-probability components, which smooths entropy variation across energy levels and reduces the sharpness of spectral contrast. 

While the peak location is largely preserved for $q \geq 1$, its shape exhibits a systematic dependence on $q$: increasing $q$ enhances the contrast between localized and extended regions in $\tilde{S}_q(E)$, resulting in a sharper and more pronounced peak in $\chi_q$, whereas smaller values of $q$ lead to broader and smoother features. This demonstrates that $q$ acts as a resolution parameter, sharpening spectral heterogeneity without altering the underlying mobility-edge physics, provided the contrast remains well defined.

Contour plots of $\tilde{S}_q(E,\lambda)$ for the AA (Fig.~\ref{plt_5}), SSH (Fig.~\ref{plt_6}), and GAA (Fig.~\ref{plt_7}) models, computed at $q=1.5$ and $L=600$, further confirm the energy-resolved transition behavior in each case. For the AA model, the line $\lambda_c = 2t$ clearly separates the high-entropic (delocalized) and low-entropic (localized) regions, demonstrating that $\tilde{S}_q$ serves as a faithful diagnostic of the global transition. Similarly, in the GAA model, the analytical mobility-edge condition (Eq.~\ref{ME_GAA}) separates high- and low-entropic regions in the $(E,\lambda)$ plane. In Fig.~\ref{plt_7}, two critical values $\lambda_{c_1}$ and $\lambda_{c_2}$ are identified, between which the mobility edge persists, consistent with the coexistence regime inferred from $\chi_q$.

The dependence of $\chi_q$ on the entropic index $q$ provides further insight into the underlying spectral structure. For $q > 1$ (subextensive regime), the Tsallis entropy becomes increasingly sensitive to high-probability components of the wavefunction, thereby emphasizing localized peaks. This enhances the contrast between localized and extended states across the spectrum, leading to a sharper and more narrowly defined peak in $\chi_q$. In contrast, for $q < 1$ (superextensive regime), the entropy assigns relatively larger weight to low-probability components, amplifying contributions from rare configurations associated with extended tails. As a result, the variation of $\tilde{S}_q(E)$ across energy becomes smoother, producing a broader and less pronounced peak in $\chi_q$.

This systematic evolution of peak height and sharpness with $q$ highlights the role of the entropic index as a tunable parameter controlling the sensitivity of the measure to different structural features of the eigenstates, and further supports the interpretation of $\chi_q$ as a probe of spectral heterogeneity. This result also indicates that excessively small values of $q$ reduce the resolution of spectral heterogeneity by overemphasizing rare-state contributions.
\subsection{Localization Physics of Different Quasiperiodic Models}

\subsubsection{Aubry--Andr\'e Model (AA): Global Localization Transition}

In the AA model, all eigenstates localize simultaneously at $\lambda_c = 2t$. Figure~\ref{plt_11}(a) shows that the IPR variance exhibits multiple peaks near $\lambda_c$, reflecting sensitivity to fluctuations of individual eigenstates. In contrast, Fig.~\ref{plt_8} and Fig.~\ref{plt_11}(b) show that $\chi_q(\lambda)$ displays only a smooth crossover without a pronounced peak. This behavior is consistent with the global nature of the transition: since all eigenstates localize collectively, the entropy profile $\tilde{S_q}(E)$ remains homogeneous across the spectrum, and its derivative does not develop singular features. Thus, entropy-gradient susceptibility captures the redistribution of information but does not yield sharply localized critical signatures in globally transitioning systems.

\subsubsection{SSH Model with Quasiperiodic Potential}
Figure~\ref{plt_9} shows that $\chi_q(\lambda)$ for the SSH model develops a pronounced peak that sharpens with increasing system size, reflecting enhanced entropy fluctuations due to the coexistence of localized and extended states. As seen in Fig.~\ref{plt_6}, localized and extended eigenstates coexist over a finite parameter range, producing a curved boundary in the energy–potential plane that signals the presence of a mobility edge.

\subsubsection{Generalized Aubry--Andr\'e Model (GAA): Mobility-Edge Scaling}
In Fig.~\ref{plt_10}, we see that the susceptibility peak persists between the previously identified critical $\lambda$ values: $\lambda_{c_1}$ and $\lambda_{c_2}$ (see Fig.~\ref{plt_7}), which confirms that the peak actually indicates the presence of a mobility edge. The interval [$\lambda_{c_1}$, $\lambda_{c_2}$] defines the regime of maximal spectral heterogeneity, consistent with the coexistence of localized and extended states.
In the GAA model, localized and extended eigenstates coexist over a finite energy window. Figure~\ref{plt_12}(a) shows that the $IPR$ variance increases broadly with $\lambda$ and exhibits strong finite-size dependence, reflecting sensitivity to rare states. In contrast, Fig.~\ref{plt_12}(b) shows that $\chi_q(\lambda)$ develops a sharp, size-stable peak at intermediate $\lambda$, with curves for larger $L$ collapsing onto one another. This indicates robust thermodynamic behavior and highlights the superiority of entropy-gradient susceptibility over $IPR$ variance in diagnosing mobility edges.

The susceptibility enhancement reflects strong energy-space inhomogeneity in $\tilde{S_q}(E)$, arising when localized and extended states occupy distinct energy regions.
\subsection{Scaling of Entropy-Gradient Susceptibility with System Size}

To assess the robustness of the entropy-gradient susceptibility against finite-size effects, 
we computed $\chi_{q}(\lambda)$ for the generalized Aubry--André (GAA) model at system sizes 
$L=600, 800,$ and $1000$ at $q=1.5$. The results are shown in Fig.~\ref{plt_ex1}. 

Across all sizes, the susceptibility develops a pronounced peak at intermediate values of the quasiperiodic potential strength $\lambda$, signaling the coexistence of localized and extended eigenstates characteristic of mobility-edge physics. Importantly, while the peak height increases 
and sharpens with larger $L$, its position remains nearly invariant. This collapse of curves onto a common peak location demonstrates that the entropy-gradient susceptibility provides a size-stable diagnostic of spectral heterogeneity. 

In contrast to conventional measures such as the inverse participation ratio (IPR) variance, which exhibit strong finite-size dependence and sensitivity to rare states, the entropy-gradient susceptibility captures collective spectral fluctuations in a manner that is both robust and universal. The sharpening of the peak with increasing $L$ reflects enhanced resolution of the 
mobility edge in larger systems, while the stability of its position confirms that the diagnostic faithfully tracks the analytic mobility-edge condition of the GAA model. These results highlight the superiority of the entropy-gradient framework in distinguishing global localization transitions from mobility-edge scenarios, offering a reliable probe of energy-resolved localization that remains valid in the thermodynamic limit.
\par
\begin{figure}
    \centering
    \includegraphics[width=0.48\textwidth]{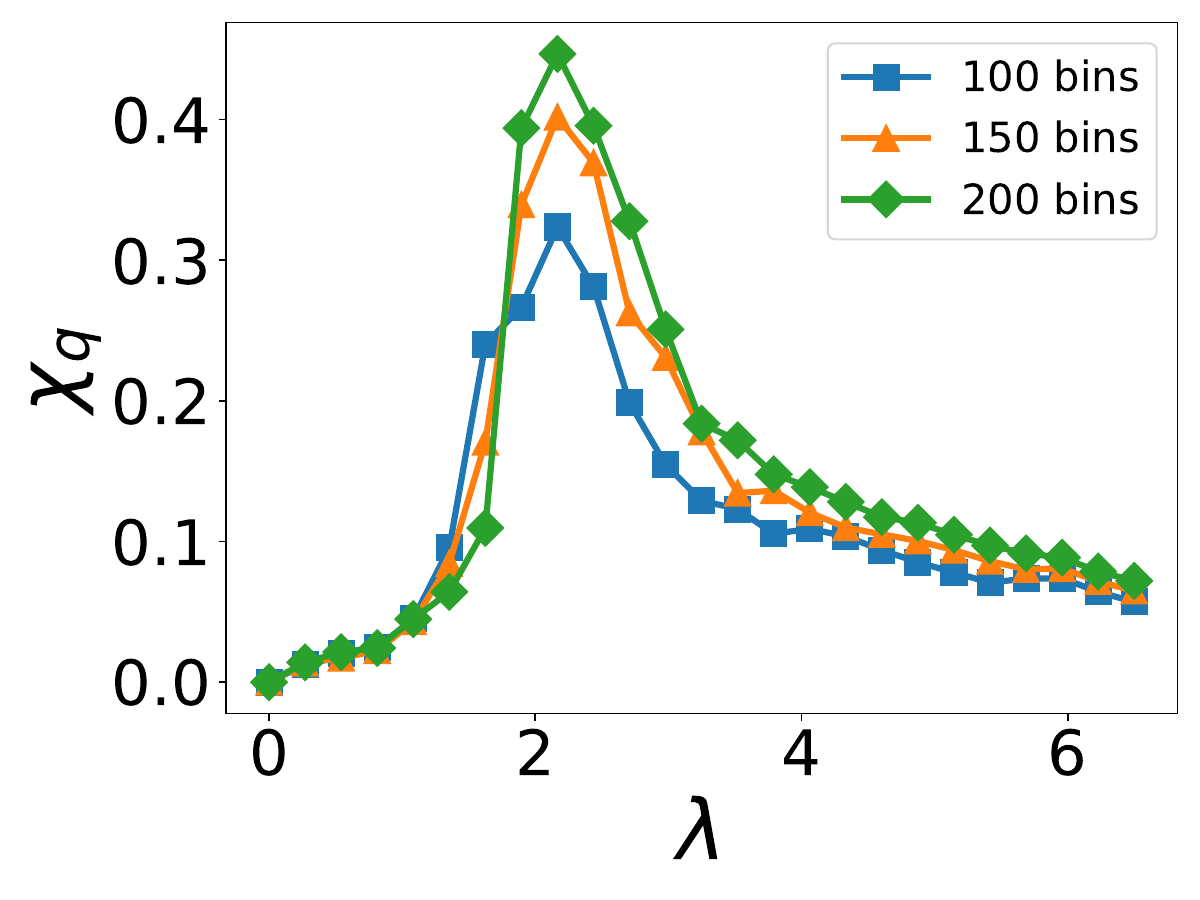}
    \caption{Entropy-gradient susceptibility $\chi_q(\lambda)$ of the GAA model as a function of the quasiperiodic potential strength $\lambda$ for $q=1.5$ for different bin sizes $100, 150, 200$. The peak position remains unchanged, and the peak becomes sharper with increasing bin size.}
    \label{plt_ex2}
\end{figure}

 We also plot in Fig.~\ref{plt_ex2} the entropy-gradient susceptibility $\chi_q(\lambda)$ of the GAA model as a function of the quasiperiodic potential strength $\lambda$ for $q=1.5$ for different bin sizes $100, 150, 200$. The peak position remains unchanged for $q >1$, and the peak becomes sharper with increasing bin size. These demonstrate the robustness of the peak to understand the mobility edge.
\section{Analytical Insights from Limiting Cases}\label{sec_4}

Although localization in quasiperiodic systems is typically explored numerically in the light of information-theoretic measures, several controlled analytical considerations elucidate the behavior of the entropy-gradient susceptibility introduced above. In particular, simple estimates in limiting regimes clarify why gradient-based measures sharply distinguish between global-localization transitions and mobility-edge scenarios.
\subsection{Extended-State Limit}

In the fully extended regime, eigenstates are approximately uniform over the lattice,
\begin{equation}
p_i \simeq \frac{1}{N},
\end{equation}
where $N$ is the total number of lattice sites. Substituting this distribution into the Tsallis entropy yields
\begin{equation}
\sum_i p_i^q = N \left(\frac{1}{N}\right)^q = N^{1-q},
\end{equation}
and therefore
\begin{equation}
S_q = \frac{1 - N^{1-q}}{q-1} \equiv S_q^{\mathrm{max}}.
\end{equation}
The normalized entropy $\tilde S_q = S_q / S_q^{\mathrm{max}}$ thus approaches unity for all extended eigenstates, independently of energy. Consequently, the entropy profile across the spectrum is flat, and its gradient with respect to energy vanishes:
\begin{equation}
\frac{d\tilde S_q}{dE} \approx 0 \quad \Rightarrow \quad \chi_q \to 0 ~(\text{delocalized regime}).
\end{equation}
This explains the small susceptibility values observed in the weak-modulation regime, or particularly in the delocalized region.

\subsection{Strong Localization Limit}

In the opposite limit of strong quasiperiodic modulation, eigenstates become exponentially localized with localization length $\xi \ll N$. The representative probability distribution may be approximated as
\begin{equation}
p_i \sim \frac{1}{\xi} e^{-|i-i_0|/\xi}.
\end{equation}
The effective support of such a state scales as $N_{\mathrm{eff}} \sim \xi$, leading to
\begin{equation}
\sum_i p_i^q \sim \xi^{1-q},
\end{equation}
and hence
\begin{equation}
S_q \sim \frac{1 - \xi^{1-q}}{q-1}.
\end{equation}
After normalization, the entropy behaves as
\begin{equation}
\tilde S_q \sim \left( \frac{\xi}{N} \right)^{1-q},
\end{equation}
which is small for $\xi \ll N$. Importantly, when all eigenstates are localized with comparable localization lengths, entropy values cluster narrowly across the spectrum. The entropy profile is nearly constant in energy, so its gradient remains finite but uniform:
\begin{equation}
\chi_q \to \text{saturates (localized regime)}.
\end{equation}

\subsection{Mobility-Edge Regime}
\label{subsec:susceptibility_derivation}

We now present a detailed analytical discussion of the scaling form of the entropy-gradient susceptibility in the presence of a mobility edge. The derivation relies only on coarse-graining in energy and does not assume any microscopic details of the eigenstates.

\subsubsection{Definition of susceptibility.}
The entropy-gradient susceptibility is defined as the coarse-grained average of the spectral entropy slope, as defined in Eq.~\ref{susceptibility}, where the average is taken over energy bins across the spectrum.

\subsubsection{Two-population approximation.}
In the mobility-edge regime, the spectrum separates into two statistically distinct classes of eigenstates:
\begin{itemize}
\item a fraction $(1-f)$ of extended states with normalized entropy $\tilde S_q = S_E \approx 1$,
\item a fraction $f$ of localized states with normalized entropy $\tilde S_q = S_L \ll 1$.
\end{itemize}
 If the normalized entropy of the $i^{th}$ extended state can be written as 
\begin{equation}
    S_E^i=S_E +\delta_{E_i},
\end{equation}
where $\delta_{E_i}$ is the fluctuation term arising from the spectral fluctuation. Within this approximation, we consider that entropy fluctuations within each class are negligible compared to the sharp change between $S_E$ and $S_L$. Thus, 
\begin{equation}
    S_E^i=S_E, \forall i.
\end{equation}
\subsubsection{Entropy profile.}
The entropy as a function of energy can be modeled as a step-like function,
\begin{equation}
\tilde S_q(E) \approx
\begin{cases}
S_E & E < E_c , \\
S_L & E > E_c ,
\end{cases}
\end{equation}
where $E_c$ denotes the mobility edge. Upon coarse-graining over an energy window $\Delta E$, the gradient is approximated as
\begin{equation}
\frac{d\tilde S_q}{dE} \sim \frac{S_E - S_L}{\Delta E}.
\end{equation}

\subsubsection{Scaling of susceptibility.}

We approximate the spectrum as consisting of two distinct classes of eigenstates: extended states with normalized Tsallis entropy $S_E$ and localized states with entropy $S_L$. Let the fraction of extended states below energy $E$ be denoted by $f(E)$. The coarse-grained entropy profile can then be written as
\begin{equation}
\tilde{S_q}(E) \;\approx\; f(E)\,S_E + \big(1-f(E)\big)\,S_L.
\end{equation}

Differentiating with respect to energy yields
\begin{equation}
\frac{d\tilde{S_q}(E)}{dE} \;=\; \frac{df(E)}{dE}\,(S_E - S_L).
\end{equation}

The entropy-gradient susceptibility, defined as the average of this derivative across the spectrum, is therefore controlled by the rate of change of the extended-state fraction. If the transition between extended and localized states occurs across an energy window of width $\Delta E$, a natural choice for $f(E)$ can be taken from FD (Fermi Dirac) statistics~\cite{beale2021statistical} as follows,
\begin{equation}
f(E) \;\approx\; \frac{1}{1 + e^{(E-E_c)/\Delta E}},
\end{equation} (where the thermal energy $k_B T$ is approximated as $\Delta E$)
whose derivative is
\begin{equation}
\frac{df(E)}{dE} \;\sim\; \frac{1}{\Delta E}\,f(E)\big(1-f(E)\big).
\end{equation}

Substituting back, the typical gradient magnitude becomes
\begin{equation}
\frac{d\tilde{S_q}(E)}{dE} \;\sim\; \frac{1}{\Delta E}\,f(E)\big(1-f(E)\big)\,(S_E - S_L).
\end{equation}

Averaging over the transition region then gives the compact form
\begin{equation}
\chi_q \;\propto\; f(1-f)\,\frac{|S_E - S_L|}{\Delta E},
\label{eq:chi_grad_final}
\end{equation}
which is Eq.~(\ref{eq:chi_grad_final}). This expression highlights three essential ingredients: (i) the coexistence factor $f(1-f)$, maximal when localized and extended states are both present; (ii) the entropy contrast $|S_E - S_L|$ between the two sectors; and (iii) the energy width $\Delta E$ of the transition region. Together, these determine the strength of the entropy-gradient susceptibility as a probe of spectral heterogeneity.

Equation~(\ref{eq:chi_grad_final}) demonstrates that the susceptibility is maximized when localized and extended states coexist in comparable proportions ($f = 1/2$), and when the entropy contrast $|S_E - S_L|$ is sharp. In contrast, when the spectrum is homogeneous ($f=0$ or $f=1$), the entropy profile is flat, and the susceptibility vanishes. This explains why mobility-edge systems exhibit a pronounced susceptibility peak, while models undergoing global localization transitions display only a smooth enhancement followed by saturation.

\section{Conclusion} \label{sec_5}

In this work, we have developed an information-theoretic framework based on Tsallis entropy and its spectral variation to probe localization phenomena in quasiperiodic systems. By analyzing the energy dependence of the entropy, we introduced an entropy-gradient susceptibility that quantifies variations in eigenstate structure across the spectrum and provides a direct measure of spectral heterogeneity. 

Our results demonstrate a clear distinction between global localization transitions and mobility-edge physics. In the Aubry--André model, where all eigenstates undergo a collective transition, the entropy varies smoothly, resulting in only a broad crossover in the proposed measure. In contrast, in systems hosting mobility edges---including the quasiperiodically modulated Su--Schrieffer--Heeger chain and the generalized Aubry--André model---the coexistence of localized and extended states leads to sharp entropy variations, producing a pronounced peak in the susceptibility. This peak serves as a robust diagnostic of energy-resolved localization and enables the identification of characteristic scales associated with the onset and completion of the mobility-edge regime.

We find that the qualitative features of the entropy-gradient susceptibility persist over a broad range of the Tsallis deformation parameter $q$, with systematic variations reflecting the relative weighting of rare and dominant spectral contributions. This behavior indicates that the observed signatures arise from intrinsic spectral structure rather than from a specific choice of moment, and highlights the approach's flexibility as a tunable probe of localization properties. In this sense, the proposed measure complements conventional diagnostics such as the inverse participation ratio, while exhibiting reduced sensitivity to finite-size effects and rare-state fluctuations.

More broadly, our results establish entropy fluctuations as a physically transparent probe of spectral heterogeneity in quasiperiodic systems. The framework can be naturally extended to interacting systems, nonequilibrium dynamics, and many-body localization, where energy-resolved structure plays a central role. 

Lastly, recent advances in engineered quasiperiodic platforms provide promising routes for experimental realization in setups such as ultracold atomic systems, which offer precise control over quasiperiodic potentials and access to eigenstate distributions, while photonic waveguide arrays enable tunable disorder and direct imaging of wavefunction profiles. In such settings, spectral entropy variations can be reconstructed from measured spatial distributions, providing a practical avenue for experimentally probing mobility-edge physics through entropy-based diagnostics. These prospects underscore the broader applicability of the present framework and open the door to direct experimental tests of information-theoretic approaches to localization.
\section{Acknowledgments}
AG gratefully acknowledges Dr. Shaon Sahoo for the discussions in related topic. 
\bibliography{references}

\end{document}